\documentclass[a4paper,11pt]{article}
\usepackage{pos}

\title{Forward $\eta_Q$ production impact factor at NLO}

\author*[a]{Maxim Nefedov}

\affiliation[a]{Universit\'e Paris-Saclay, CNRS, IJCLab, 91405 Orsay, France}

\emailAdd{maxim.nefedov@desy.de}

\abstract{The High-Energy Factorisation (HEF) coefficient function or impact factor for the hadroproduction of the state ${}^1S_0^{[1]}$ of a heavy quark-antiquark pair ($Q\bar{Q}$) at forward rapidities is computed at NLO in $\alpha_s$. This impact factor is an integral part of the resummation procedure for higher-order QCD corrections to the  hadroproduction cross sections of $\eta_c$ or $\eta_b$ mesons at forward rapidities in collinear factorisation (CF), enhanced by logarithms of partonic center of mass energy. The NLO computation results of which are presented in this contribution, is an important step towards an extension of aforementioned resummation beyond the Leading Logarithmic Approximation(LLA).   }

\FullConference{31st International Workshop on Deep Inelastic Scattering (DIS2024)\\
 8–12 April 2024\\
Grenoble, France\\}

\begin{document}
\maketitle

\section{Introduction}
Recently, the importance of resummation of higher-order QCD corrections enhanced by logarithms of partonic center-of-mass energy ($\hat{s}$) in the inclusive heavy-quarkonium hadroproduction~\cite{Lansberg:2021vie} and photoproduction~\cite{Lansberg:2023kzf} computations in CF had been pointed out. The goal of the present contribution is to report the result of the computation of the NLO coefficient function (or impact factor) of HEF, which is necessary to extend such computations beyond Doubly-Logarithmic Approximation (DLA), used in Refs.~\cite{Lansberg:2021vie,Lansberg:2023kzf}. Due to the lack of space in this proceedings, only the final result of the computation will be presented, postponing the discussion of it's derivation and numerical results obtained with it for a more complete publication. To put this computation in context, one considers the quarkonium transverse momentum (${\bf p}_T$) and rapidity ($y$)-differential cross section of the process $p(P_1) + p(P_2) \to Q\bar{Q}[{}^1S_0^{[1]}](p)+X$, where $P_1^\mu = \sqrt{S} n_-^\mu/2$, $P_2^\mu = \sqrt{S} n_+^\mu/2$ (with Sudakov vectors $n_{\pm}^\mu=(1,0,0,\mp 1)^\mu$ in the $pp$ center-of-momentum frame and $p^2=M^2$), which in CF can be written as:
\begin{eqnarray}
    &&\hspace{-0.5cm} \frac{d\sigma}{dy d{\bf p}_T^2} = \frac{M^2}{S} \int\limits_0^{\eta_{\max}} d\eta \int\limits_{0}^{1} dz\  f_{i}\left(x_1,\mu_F\right) f_{j} \left(x_2,\mu_F \right) \frac{d\hat{\sigma}_{ij}(\eta,z,{\bf p}_T^2)}{dz d{\bf p}_T^2},\label{eq:CF-master}
\end{eqnarray}
where $x_1=\frac{M_T e^{y}}{\sqrt{S}z}$, $x_2= \frac{M^2 z (1+\eta)}{M_T\sqrt{S}}e^{-y}$, $M_T=\sqrt{M^2+{\bf p}_T^2}$, $f_{i}(x_{1,2},\mu_F)$ are collinear PDFs of a parton $i=q,\bar{q},g$ with four-momentum $x_{1,2}P_{1,2}^\mu$ and factorisation scale $\mu_F$, $z=(n_+\cdot p)/(x_1 n_+\cdot P_1)=p_+/(x_1 P_1^+)$, $\eta=(\hat{s}-M^2)/M^2$ with $\hat{s}=x_1x_2 S$ being the squared center-of-mass energy in the partonic subprocess of CF, $d\hat{\sigma}_{ij}$ is the partonic coefficient function of a said subprocess and $\eta_{\max}=S/M^2-1$. With the aim to resum LL ($\propto \alpha_s^n \ln^{n-1}(1+\eta)$) and NLL ($\propto \alpha_s^{n+1} \ln^{n-1}(1+\eta)$) corrections to the $d\hat{\sigma}$ in Eq.~(\ref{eq:CF-master}) at $\eta\gg 1$, one writes the following HEF-resummation formula in the gluon channel\footnote{Results for $\hat{\sigma}_{iq}$ will be presented elswere.}:
\begin{eqnarray}
   && \frac{d\hat{\sigma}_{ig}^{\text{(HEF, NLLA)}}}{dz d{\bf p}_T^2} = \frac{1}{2M^2} \int \frac{d^2{\bf q}_T}{\pi} {\cal C}_{ig} \Bigl(1+\eta,{\bf q}_T^2,\mu_F,\mu_R \Bigr) \nonumber \\ 
   && \times \left[ {\cal H}_{gg}^{\text{(LO)}}({\bf p}_T^2) \delta(1-z)\delta({\bf q}_T^2-{\bf p}_T^2) + \frac{\alpha_s(\mu_R)}{2\pi} {\cal H}_{gg}^{\text{(NLO)}}({\bf q}_T,z,{\bf p}_T) \right], \label{eq:HEF-master}
\end{eqnarray}
where the resummation factor ${\cal C}_{ig}$ in the LLA is defined e.g. in \cite{Lansberg:2021vie} and should be extended with NLLA corrections to obtain the complete NLLA result for the $d\hat{\sigma}$. The LO HEF coefficient function is: 
$   {\cal H}_{gg}^{\text{(LO)}}({\bf p}_T) =  \frac{32\pi^3\alpha_s^2}{N_c^2(N_c^2-1)} \frac{\left\langle{\cal O}\left[^1S_0^{[1]} \right] \right\rangle}{M^3} \frac{M^4}{(M^2+{\bf p}_T^2)^2}, $
with the Long-Distance Matrix Element of Non-Relativistic QCD factorisation~\cite{Bodwin:1994jh} being $\left\langle{\cal O}\left[^1S_0^{[1]} \right] \right\rangle=2N_c |R(0)|^2/(4\pi)$. The NLO correction to the HEF coefficient function in Eq.~(\ref{eq:HEF-master}) in the gluon channel can be decomposed as follows:
\begin{equation}
     {\cal H}_{gg}^{\text{(NLO)}}({\bf q}_T,z,{\bf p}_T) = {\cal H}_{gg}^{\text{(real, fin.)}}({\bf q}_T,z,{\bf p}_T) + {\cal H}_{gg}^{\text{(NLO, analyt.)}} ({\bf q}_T,z,{\bf p}_T). \label{eq:H-NLO-decomp}
\end{equation}

\section{The real-emission contribution}

The real-emission contribution to the NLO impact factor (\ref{eq:H-NLO-decomp}) in the gluon channel comes from the process $g(x_1P_1)+R_-({\bf q}_T)\to Q\bar{Q}[{}^1S_0^{[1]}](p)+g(k)$, where $R_-$ is the Reggeised gluon, the exact matrix element of this process can be written as:
\begin{equation}
\Tilde{H}_{Rg}=\frac{C_A \left(M^2+{\bf p}_T^2\right)^2}{s^2 t^2 u^2 \left(M^2-s\right)^2 \left(M^2-t+t_1\right){}^2 \left(M^2-u\right)^2}\sum\limits_{n=-1}^3 (1-z)^n w_n(s,t,u). \label{eq:H-Rg-real}
\end{equation}
where $t_1={\bf q}_T^2$, Mandelstam variables $s = M^2\rho-{\bf q}_{T}^2$, $t = -[ M^2(\rho z-1)-{\bf p}_{T}^2 ]/z$, $u = -[ M^2(1-z)+{\bf p}_{T}^2 ]/z$, $\rho=\frac{({\bf p}_T-{\bf q}_T)^2}{M^2(1-z)} + \frac{M^2+{\bf p}_T^2}{M^2 z}$, and coefficients:
\begin{eqnarray}
&& \hspace{-0.5cm} w_{-1}= s t t_1 u \Bigl( t + t_1 + u\Bigr) \Bigl( s + 2 t_1 + u\Bigr) \
\Bigl( s^2 [ t_1 - u] [ t + t_1 + u] + t [ t + t_1] [ t_1 + u] [ 2 t_1 + u] \nonumber \\ 
&& \hspace{-0.5cm} + s [ t_1^3 - t_1 u^2 \
+ t^2 ( t_1 + u) + 2 t t_1 ( 2 t_1 + u)]\Bigr), \\
 && \hspace{-0.5cm}w_0=-u \Biggl\{ -s^5 t t_1 \Bigl( t + t_1 + u\Bigr)^2 - t^2 u \Bigl( t + \
t_1\Bigr)^2 \Bigl( t + t_1 + u\Bigr)^2 \Bigl( 2 t_1 + u\Bigr)^2 - s^4 \Bigl[ t + t_1 + u\Bigr] \Bigl[ -u ( t_1 + u)^3  \nonumber \\ 
 && \hspace{-0.5cm}- u] + t ( t_1 + u)^2 [ 2 t_1 + t^3 [ 5 t_1 + u] + t^2 [ t_1 + u] [ 10 t_1 + u]\Bigr] - s t \
\Bigl[ t + t_1\Bigr] \Bigl[ t + t_1 + u\Bigr] \Bigl[ 2 t_1 + u\Bigr]  \nonumber \\ 
 && \hspace{-0.5cm}\times\Bigl[ 2 t^3 u + t_1^2 u [ \
t_1 + u] + t u [ 2 t_1 + u] [ 5 t_1 + 2 u] + 2 t^2 [ t_1 + u] [ 4 t_1 + 3 u]\Bigr]  \nonumber \\ 
 && \hspace{-0.5cm} + s^2 \Bigl[ t + t_1 + \
u\Bigr] \Bigl[ -t^5 u + t_1^2 u ( t_1 + u)^3 + t t_1 [ t_1 + u] [ 2 t_1 + u] [ t_1^2 - t_1 u + 2 u^2] - \
t^4 [ 16 t_1^2 + 21 t_1 u + 7 u^2] \nonumber \\
&& \hspace{-0.5cm} - t^3 [ 2 t_1 + u] [ 22 t_1^2 + 29 t_1 u + 8 u^2] - 2 t^2 \
t_1 [ 5 t_1^3 + 24 t_1^2 u + 19 t_1 u^2 + 4 u^3]\Bigr] \nonumber \\ 
 && \hspace{-0.5cm} + s^3 \Bigl[ 2 t_1 u ( t_1 + u)^4 - 2 t^5 [ 2 t_1 + \
u] - 2 t^2 t_1 [ t_1 + u] [ 10 t_1^2 + 22 t_1 u  + 7 u^2] - t^4 [ 32 t_1^2 + 33 t_1 u + 8 u^2] \nonumber \\
&& \hspace{-0.5cm} + t ( t_1 + u)^2 [ \
t_1^3 + 2 t_1^2 u + 7 t_1 u^2 + 2 u^3] - t^3 [ 49 t_1^3 + 101 t_1^2 u + 54 t_1 u^2 + 8 \
u^3]\Bigr]\Biggr\} , \\
&& \hspace{-0.5cm}w_1=s u \Bigl[ s + t + t_1\Bigr] \Bigl[ t + t_1 + u\Bigr] \Bigl[ s^3 \
\Bigl( t + u\Bigr) \Bigl( 2 t^2 + t t_1 - 2 ( t_1 + u)^2\Bigr) + t \Bigl( 2 t_1 + u\Bigr) \Bigl( t t_1 u [ \
-4 t_1 + u]  \nonumber \\ 
 && \hspace{-0.5cm} + 2 t^3 [ 2 t_1 + u] - t_1^2 u [ t_1 + 2 u] + 2 t^2 [ -4 t_1^2 + t_1 u + u^2]\Bigr) + \
s^2 \Bigl( 2 t^4 - 4 t_1 u ( t_1 + u)^2 + 2 t^3 [ 3 t_1 + 5 u] \nonumber \\ 
 && \hspace{-0.5cm} + t^2 [ -2 t_1^2 + t_1 u + 6 u^2] - \
t [ 5 t_1^3  + 14 t_1^2 u + 14 t_1 u^2 + 2 u^3]\Bigr) \nonumber \\ 
 && \hspace{-0.5cm}  + s \Bigl( -2 t_1^2 u ( \
t_1 + u)^2 + 4 t^4 [ 2 t_1 + u] + t^3 [ -4 t_1^2 + 20 t_1 u + 10 u^2] + t^2 [ -8 t_1^3 - 18 t_1^2 \
u + t_1 u^2 + 2 u^3] \nonumber \\
&& \hspace{-0.5cm} - t t_1 [ 2 t_1^3 + 10 t_1^2 u + 14 t_1 u^2 + 3 \
u^3]\Bigr)\Bigr] \\
&& \hspace{-0.5cm} w_2 = s \Biggl\{ s^5 \Bigl( t + t_1 + u\Bigr)^2 \Bigl[ t - u\Bigr] \Bigl[ \
t + u\Bigr] + s^4 \Bigl[ t + t_1 + u\Bigr] \Bigl[ -2 u^3 [ t + 2 t_1] + 2 t^2 [ t + t_1] [ t + 3 \
t_1]  \nonumber \\ 
 && \hspace{-0.5cm} + u^2 [ 2 t^2 - 5 t t_1 - 4 t_1^2] + t u [ 6 t^2 + 9 t t_1 + t_1^2]\Bigr] + s^2 \Bigl[ t \
t_1 u^5 + 4 t^2 t_1 ( t + t_1)^3 [ t + 3 t_1] \nonumber \\ 
 && \hspace{-0.5cm} + 2 u^4 [ t^3 + 6 t^2 t_1 - 2 t_1^3] + 2 u^3 [ 4 t^4 + 27 t^3 t_1 \
+ 29 t^2 t_1^2 - 6 t t_1^3 - 4 t_1^4]  \nonumber \\ 
 && \hspace{-0.5cm} + t u [ t + t_1] [ 2 t^4 + 32 t^3 t_1 + 90 t^2 t_1^2 + 51 t t_1^3 \
+ 5 t_1^4] + u^2 [ 8 t^5 + 73 t^4 t_1 + 163 t^3 t_1^2 + 84 t^2 t_1^3 - 6 t t_1^4 - 4 t_1^5]\Bigr] \nonumber \\ 
 && \hspace{-0.5cm} + s \
\Bigl[ 4 t^2 t_1^2 ( t + t_1)^4 + 2 t t_1 u^5 [ t + t_1]  + u^4 [ t^4 + 17 t^3 t_1 + 24 t^2 t_1^2 + 4 t t_1^3 - t_1^4]  \nonumber \\ 
 && \hspace{-0.5cm} + 2 t \
t_1 u [ t + t_1] [ 2 t^4 + 20 t^3 t_1 + 36 t^2 t_1^2 + 11 t t_1^3 + t_1^4] + u^3 [ 2 t^5 + 33 t^4 t_1 + \
101 t^3 t_1^2 + 64 t^2 t_1^3 - 4 t t_1^4  - 2 t_1^5] \nonumber \\
\end{eqnarray}
\begin{eqnarray}
&& \hspace{-0.5cm} + u^2 [ t^6 + 22 t^5 t_1 + 117 t^4 t_1^2 + 172 t^3 \
t_1^3 + 57 t^2 t_1^4 - 4 t t_1^5 - t_1^6]\Bigr] + t t_1 u \Bigl[ t + t_1\Bigr] \Bigl[ t + t_1 + u\Bigr] \Bigl[ 2 \
t_1 + u\Bigr] \nonumber \\
&& \hspace{-0.5cm} \times \Bigl[ t_1 u [ -t_1 + u] + 4 t^2 [ 2 t_1 + u] + t u [ 6 t_1 + u]\Bigr] + s^3 \Bigl[ t^6 - 6 t_1^2 u^2 ( t_1 \
+ u)^2 + 4 t^5 [ 3 t_1 + 2 u] \nonumber \\
&& \hspace{-0.5cm} + 2 t t_1 u [ t_1 + u] [ 2 t_1^2 - 7 t_1 u - 2 u^2] + t^4 [ 34 \
t_1^2 + 53 t_1 u + 15 u^2] + t^3 [ 36 t_1^3 + 87 t_1^2 u + 61 t_1 u^2 + 8 u^3]  \nonumber \\ 
 && \hspace{-0.5cm} + t^2 t_1 [ 13 t_1^3 + 46 t_1^2 \
u + 41 t_1 u^2 + 14 u^3]\Bigr]\Biggr\},\\
&& \hspace{-0.5cm}w_3 = s t t_1 u \Bigl( s + t + t_1\Bigr) \Bigl( s + 2 t_1 + u\Bigr) \
\Bigl( s^2 [ t - u] [ t + t_1 + u] + t_1 [ t + t_1 + u] [ 2 t t_1 + u ( t + t_1)]  \nonumber \\ 
 && \hspace{-0.5cm} + s t [ ( 3 \
t_1 - u) ( t_1 + u) + t ( 3 t_1 + u)]\Bigr).
\end{eqnarray}

The integral of the Eq.~(\ref{eq:H-Rg-real}) over phase space of emitted gluon ($g(k)$) is divergent at ${\bf k}_T\to 0$ due to soft and collinear singularities and at $z\to 1$ due to the rapidity divergence. In the contribution ${\cal H}_{gg}^{\text{(real, fin.)}}$  to the NLO impact factor (\ref{eq:H-NLO-decomp}): 
\begin{equation}
    {\cal H}_{gg}^{\text{(real, fin.)}} ({\bf q}_T^2,z,{\bf p}_T^2)=  {\cal H}_{gg}^{\text{(LO)}}({\bf p}_T^2) \int\limits_0^{2\pi}\frac{d\phi}{2\pi} \left[ \frac{\Tilde{H}_{Rg}({\bf q}_T,z,{\bf p}_T) }{z(1-z) {\bf q}_T^2} - {\cal J}^{\text{(sub.)}}_{Rg}({\bf q}_T, z,{\bf p}_T,0) \right], \label{eq:Hgg-subtr}
\end{equation}
these divergences are subtracted by the following local subtraction term, similar to one used in Ref.~\cite{Hentschinski:2020tbi}:
\begin{equation}
    {\cal J}^{\text{(sub.)}}_{Rg}({\bf q}_T,z,{\bf p}_T,r) = \frac{2C_A}{{\bf k}_T^2} \left[ \frac{1-z}{(1-z)^2 + r\frac{{\bf k}_T^2}{(x_1P_1^+)^2}} + \Delta p_{gg}({\bf q}_T,z,{\bf p}_T) \right], \label{eq:J-subtr}
\end{equation}
where $\Delta p_{gg}({\bf q}_T,z,{\bf p}_T)=z(1-z) + 2[{\bf k}_T^2 {\bf q}_T^2 - ({\bf k}_T {\bf q}_T)^2]/(z {\bf k}_T^2 {\bf q}_T^2) - [3{\bf k}_T^2 {\bf q}_T^2 - 2({\bf k}_T {\bf q}_T)^2]/({\bf k}_T^2 {\bf q}_T^2)$, ${\bf k}_T={\bf q}_T-{\bf p}_T$ and $r\ll 1$ is the regularisation parameter for rapidity divergence at $z\to 1$, which is set to zero in Eq.~(\ref{eq:Hgg-subtr}). In this way the Eq. (\ref{eq:Hgg-subtr}) can be safely integrated over $z$ with the PDF in Eq.~(\ref{eq:CF-master}) and over ${\bf q}_T$ with the ${\cal C}_{gg}$  in Eq.~(\ref{eq:HEF-master}) in two dimensions for ${\bf q}_T$ and at $r=0$, while the integral of the subtraction term (\ref{eq:J-subtr}) in dimensional regularisation and at $0< r\ll 1$ had been computed analytically and included into the contribution described in the next section. 

\section{One-loop contribution and integrated subtraction term}

One-loop correction to the amplitude $g(x_1P_1)+R_-({\bf q}_T)\to Q\bar{Q}[{}^1S_0^{[8]}]$ is computed in Ref.~\cite{Nefedov:2024swu} using the same ``tilted-Wilson-line'' regularisation for rapidity divergence as in Eq.~(\ref{eq:J-subtr}). The one-loop correction is combined with: i) integral of Eq.~(\ref{eq:J-subtr}) over phase-space of a gluon ($g(k)$); ii) the ultraviolet counterterms; iii) the transition terms to the regularisation scheme for rapidity divergence via the cutoff in the target light-cone momentum component $k_-$ (``the HEF scheme'', Sec.~6.2 of Ref.~\cite{Nefedov:2024swu}). After that, all IR, UV and rapidity divergences ($\sim \ln r$) cancel with an exception of the collinear divergence, proportional to the DGLAP splitting function $P_{gg}(z)$, which is absorbed into the collinear PDF of the projectile gluon $g(x_1 P_1)$. After these cancellations, the following finite remainder is left, which contributes to the Eq.~(\ref{eq:H-NLO-decomp}):
\begin{eqnarray*}
 &&  {\cal H}_{gg}^{\text{(NLO, analyt.)}} ({\bf q}_T,z,{\bf p}_T) = {\cal H}_{gg}^{\text{(LO)}}({\bf p}^2_T) \int d^2 {\bf k}_T K_{\rm BFKL}({\bf q}_T,{\bf k}_T,{\bf p}_T) \nonumber \\
 && \times\left[ \frac{1}{(1-z)_+} + \Delta p_{gg}({\bf q}_T,z,{\bf p}_T) + \delta(1-z)\ln\left(\frac{M^2+{\bf p}_T^2}{{\bf k}_T^2}\right) \right] \\
 && +  {\cal H}_{gg}^{\text{(LO)}}({\bf p}^2_T) \delta({\bf q}_T^2-{\bf p}_T^2) \left\{ -\ln\frac{\mu_F^2}{{\bf p}_T^2} P_{gg}(z) + 2C_A \frac{1-z}{z} \right. \nonumber \\
 &&\left. +\delta(1-z) \left[ -\frac{\pi^2}{2}C_A +\frac{8}{3}C_A - \frac{5}{3}\beta_0 - 2C_F \left( 2 + \frac{3}{2} \ln \frac{4{\bf p}_T^2}{M^2} \right)  +\beta_0 \ln \frac{\mu_R^2}{{\bf p}_T^2} + F_{{}^1S_0^{[1]}}({\bf p}_T^2/M^2) \right] \right\} ,
\end{eqnarray*}
where $K_{\rm BFKL}=\frac{2 C_A}{{\bf k}_T^2}\left[\delta^{(2)}({\bf q}_T-{\bf k}_T-{\bf p}_T) - \frac{{\bf p}_T^2}{{\bf p}_T^2 + {\bf k}_T^2} \delta^{(2)}({\bf q}_T-{\bf p}_T)\right]$, $\beta_0=11C_A/3-2n^{\text{(massless)}}_F/3$. The function $ F_{{}^1S_0^{[1]}}(\tau)=C_F  C[gR\to {}^1S_0^{[1]},C_F] + C_A C[gR\to {}^1S_0^{[1]},C_A]$ is the finite part of the one-loop impact factor correction, computed in the Ref.~\cite{Nefedov:2024swu}, Eqns.~(5.13) -- (5.17).

{\bf Acknowledgments:} This project has received funding from the European Union's Horizon 2020
research and innovation programme under grant agreement No.~101065263
for the Marie Sk{\l}odowska-Curie action ``RadCor4HEF'', under grant
agreement No.~824093 in order to contribute to the EU Virtual Access
{\sc NLOAccess} and to the JRA Fixed-Target Experiments at the LHC.
 This project has also received funding from the Agence Nationale de la
Recherche (ANR) via the grant ANR-20-CE31-0015 (``PrecisOnium'') and via
the IDEX Paris-Saclay ``Investissements d’Avenir'' (ANR-11-IDEX-0003-01)
through the GLUODYNAMICS project funded by the ``P2IO LabEx (ANR-10-LABX-0038)''.
This work  was also partly supported by the French CNRS via the IN2P3 projects ``GLUE@NLO'' and ``QCDFactorisation@NLO'' as well as via the COPIN-IN2P3 project \#12-147 ``$k_T$ factorisation and quarkonium production in the LHC era''.

\end{document}